\documentclass[aps,twocolumn,prl,showpacs,showkeys,floatfix]{revtex4} 
\usepackage{epsfig} 
\usepackage{makeidx} \makeindex 
\usepackage{amsmath,amssymb,amsthm} 

\begin{document}

\title{Structured environments in solid state systems: crossover from Gaussian to non-Gaussian behavior}

\author{E. Paladino$^{(1)}$, A. G. Maugeri$^{(1)}$, M. Sassetti $^{(2)}$, G. Falci $^{(1)}$ 
 and U. Weiss $^{(3)}$}

\affiliation{ 
$^{(1)}$ MATIS CNR-INFM, Catania  \& 
Dipartimento di Metodologie Fisiche e Chimiche, 
Universit\'a di Catania,
95125 Catania, Italy.\\ 
$^{(2)}$ Dipartimento di Fisica, Universit\`a di Genova \& LAMIA CNR-INFM,
16146 Genova, Italy. \\
$^{(3)}$ II. Institut f\"ur Theoretische Physik,  
Universit\"at Stuttgart, D-70550 Stuttgart, Germany. 
}

\begin{abstract}
The variety of noise sources typical of the solid state represents the main limitation toward the 
realization of controllable and reliable quantum nanocircuits, as those allowing quantum computation. 
Such ``structured environments'' are characterized by a  non-monotonous noise spectrum sometimes
showing resonances at selected frequencies. 
Here we focus on a prototype structured environment model: 
a two-state impurity linearly coupled to a dissipative harmonic bath.
We identify the time scale separating Gaussian and non-Gaussian 
dynamical regimes of the Spin-Boson impurity.
By using a path-integral approach we show that a qubit interacting
with such a structured bath  may probe the variety of environmental dynamical regimes.
\end{abstract}


\date{\today} 
\maketitle 

\section{Introduction}

Controlled coherent dynamics of solid state devices has been demonstrated in recent 
years~\cite{nature-nak,kn:exp}. 
Compared to other implementations, solid state qubits suffer from stronger 
broadband noise originating from sources with different character.
The main limitation toward the realization of controllable and reliable
quantum circuits allowing quantum computation is decoherence due to material 
(and device) dependent noise sources. The resulting noise spectrum is 
non-monotonous and sometimes characterized by resonances.
Often these features may be attributed to interaction with a
nonlinear and  non-Markovian environment~\cite{PRL02}.

In superconducting nanocircuits a particularly detrimental role is 
played by fluctuating impurities located in the insulating material 
surrounding the qubit, which are responsible for charge noise and flux 
noise~\cite{kn:nak02,kn:ithier}.
Background charges are known to be responsible for low-frequency
$1/f$ noise~\cite{kn:zorin}, moreover experiments with Josepshon devices 
suggested that spurious two-level systems may also affect 
high-frequency noise~\cite{kn:martinis}. Connections between low and
high-frequency noise features have been suggested in the recent 
experiment Ref.~\cite{astafiev04}.
Different microscopic mechanisms~\cite{faoro05} and effective 
models~\cite{shnirman05} have been recently proposed to explain the observed 
spectral features. 

Predicting decoherence originating from such a structured 
environment often responsible for non-Gaussian noise is a non trivial task, which
has attracted a lot of attention in the past years 
\cite{PRL02,kn:bruder,Shnirman03,Altshuler,PRL05}.
A well established scheme consists in 
studying the reduced dynamics of an extended system composed of the qubit
and of the environmental degrees of freedom responsible for non-Gaussian behavior. 
This strategy, combined with an appropriate classification of the noise sources 
(i.e. adiabatic or quantum noise), each treated via appropriate approximate tools,
provides a general scheme to deal with the variety of noise sources
typical of the solid state \cite{PRL05,chemphys06}. 
In this paper we focus on a prototype impurity model, a two-state impurity
linearly coupled to a dissipative harmonic bath.
Such Spin-Boson models have been thoroughly investigated with a variety of methods
since the '80s~\cite{leggett,book}. 
Thus the impurity dynamics is well known in a wide region of
parameters space.
This allows the identification of the impurity 
characteristic time scales and, therefore, of the conditions
where deviations from Gaussian and/or weak coupling regimes are expected.
Having this information at our disposal we will apply standard techniques
developed for quantum dissipative systems to find the qubit dynamics in the
presence of this structured bath.
The analysis will provide evidence for the appearance of non-Gaussian
effects. 
In particular, their onset 
will be shown to be related to clearly identifiable effects in the qubit behavior.

In Section 2 we introduce the qubit-impurity model and identify 
the relevant impurity dynamical quantities. In Section 3 we review the
equilibrium correlation function for a Spin-Boson impurity and identify its
correlation time. In Section 4 we study the qubit dynamics in the
presence of the Spin-Boson impurity within a path-integral approach
and discuss the main features of the crossover from weak to strong coupling.

\section{Model and relevant dynamical quantities}
\label{sec:model}
To  be specific we shall refer to superconducting qubits based on the Cooper-pair box
 \cite{nature-nak,rmp}.
Under proper conditions the device behaves as a two-state system
described in terms of Pauli matrices by ($\hbar=1$) 
\begin{equation}
{ \mathcal H}_{qubit} \,= \, -  \frac{E_C}{2} \, \sigma_z -  \frac{E_J}{ 2} \, \sigma_x \,.
\end{equation}
The charging energy $E_C$ gives the additional cost of adding an extra Cooper pair
to the superconducting island and the possibility of coherent transfer of pairs
through the junction is given by the Josephson term $E_J \, \sigma_x /2$.
The charge on the superconducting island may fluctuate because of interaction with 
uncontrolled impurities. Here we model a single impurity with a Spin-Boson model,
the overall Hamiltonian being given by  
\begin{eqnarray}
\label{eq:hamiltonian}
&& {\mathcal H} \,=\, { \mathcal H}_{qubit} +  { \mathcal H}_{SB} - {v \over 2} \, \sigma_z\,\tau_z \\
&&{ \mathcal H}_{SB}  \,= \,-  {\varepsilon \over 2} \, \tau_z
-  {\Delta \over 2} \, \tau_x \, - {1 \over 2} \,X\,\tau_z + {\mathcal H}_E \,.
\end{eqnarray}
The two-level-system impurity ($\vec{\tau}$) is coupled to a harmonic bath, described 
by $ {\mathcal H}_E = \sum_\alpha  \omega_\alpha  a^\dagger_\alpha a_\alpha $, via the collective coordinate $X$. 
Its effect on the impurity depends only on the spectral 
density $G(\omega)$ or equivalently on the power spectrum
\begin{eqnarray}
\label{eq:powerspec}
S(\omega) \,&=&\, \int_{-\infty}^{\infty} 
\, dt \, \frac{1}{2} \,
\langle {X}(t) {X}(0) + {X}(0) {X}(t) 
\rangle \, e^{i \omega t} = \nonumber \\
\,&=&\,
\pi \, G(|\omega|) \, \coth {\beta |\omega| \over 2} 
\,, 
\end{eqnarray}
where $\langle ... \rangle$ denotes the thermal average with respect to $ {\mathcal H}_E$,
and $\beta = 1/ k_B T$.
We consider the standard case when the coupling  operator is a collective displacement 
$X = \sum_\alpha \lambda_\alpha (a_\alpha + a^\dagger_\alpha)$ with ohmic 
spectral density
\begin{equation}
\label{eq:dens-spec-bosons}
G(\omega) \,=\, \sum_\alpha \lambda{}_\alpha^2 \delta(\omega - \omega_\alpha) \,=\,
2 \, K \, |\omega| \,  e^{- |\omega|/ \omega_c}  \, ,
\end{equation}
where $\omega_c$ represents the high frequency cut-off 
of the harmonic modes.

A first step in understanding the effects of damping is to view the impurity $\vec \tau$  {\em and} 
the ohmic bath as an environment for the qubit $\vec \sigma$. 
This environment is in general non-Gaussian and non-Markovian.
A Gaussian approximation of this structured bath amounts to replace it with 
an effective harmonic model directly coupled to $\sigma_z$ and with power spectrum $S_\tau(\omega)$
\begin{equation}
\label{eq:powerspec-tau}
S_\tau (\omega)\! =\! \frac{1}{2} \int_{-\infty}^\infty \hskip-8pt dt 
 \Big(\langle \tau_z(t)   \tau_z(0) + \tau_z(0) \tau_z(t) \rangle 
- \langle \tau_z \rangle_\infty^2 \Big)
\mathrm{e}^{i \omega t}, 
\end{equation}
the Fourier transform of  {\em equilibrium} symmetrized auto-correlation function of the 
impurity observable which directly couples to the qubit.
Here the thermal average is performed with respect to ${\mathcal H}_{SB}$ and
$\langle \tau_z \rangle_\infty$ is the thermal equilibrium value for $ \tau_z$.
Under this approximation and using a
master equation approach, the   relaxation and dephasing rates for the qubit $\vec \sigma$
in lowest order in the coupling $v$ read~\cite{book,kn:cohen}, 
\begin{eqnarray}
{1 \over T_1}\, &=& \, \Big({E_J \over E}\Big)^2 \,{v^2 S_\tau(E) \over 2} 
\label{relaxation}\\ 
{1 \over T_2} \,&=& \, {1 \over2 T_1} \,+ \,{1 \over T_2^*} \,=  \nonumber\\
&=&\, \Big({E_J \over E}\Big)_{}^2\frac{ v^2 S_\tau(E)}{4} + \Big({E_C \over E}\Big)^2 {v^2 S_\tau(0) \over 2}
\label{dephasing}
\end{eqnarray}
\begin{figure}[t!]
\centering
\includegraphics[width=0.38\textwidth,height=0.28\textwidth]{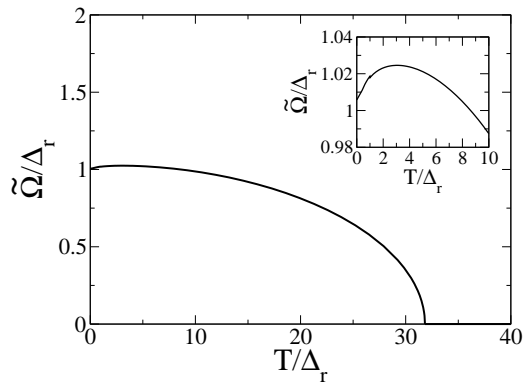}
\caption{$\tilde\Omega(T)$ of Eqs.(\ref{eq:omega1}) and (\ref{eq:omega2})
as a function of  temperature. Inset: non-monotonous behavior for $T \ll T^*=31 \Delta_r$.
Parameters are $K=0.01$, $\omega_c/\Delta_r = 31$, and we fixed $k_B=1$.
}
\label{fig:omega}
\end{figure}
where $E= \sqrt {E_C^2 +E_J^2}$ is the qubit splitting.
The validity of  this standard approach is limited to couplings  $v  \ll 1/\tau_c$~\cite{kn:cohen},
where $\tau_c$ is the range of the correlation function in Eq.(\ref{eq:powerspec-tau}) (to be defined
in the next Section).
Clearly if the impurity $\tau$ has a slow dynamics $v \, \tau_c \gg 1$, 
this picture does not apply and we have to resort to other methods. 
However previous studies on a similar model have shown that the Gaussian 
approximation may give good results even for $v \tau_c > 1$ but for shorter and shorter times, 
as long as $\tau_c$ increases~\cite{PRL02}. 

>From a different perspective the failure of the Gaussian approximation can be understood by 
viewing the qubit $\vec \sigma$ as a measuring device~\cite{kn:devoret-schoelkopf} 
for the mesoscopic system described by the Spin-Boson model involving $\vec \tau$. A rather rough 
measurement protocol (short times, averaging of results) makes the dynamics of $\vec{\sigma}$ 
essentially sensitive only to $S_\tau (\omega)$, whereas if the Spin-Boson has a slow dynamics the spin 
$\vec{\sigma}$ is able to detect also details of the dynamics of $\vec \tau$ which go 
beyond $S_\tau (\omega)$, and have to be described with more careful methods.

In the following  we will treat the impurity $\vec \tau$ on the same footing 
as the qubit $\vec \sigma$, we will apply standard methods to trace out the bosonic degrees of
freedom without any approximation on the qubit-impurity coupling.

\section{Impurity dynamics and correlation time}
\label{sec:impurity}

In this Section we will identify the characteristic time scale of
the dynamics of the {\em equilibrium} fluctuations of the Spin-Boson
impurity described by  Eq.(\ref{eq:powerspec-tau}). 
The uncoupled (i.e. for $v=0$) impurity dynamics strictly depends on
the damping strenght $K$ and on the temperature~\cite{book}. 
We consider the small damping $K \ll 1$ regime where 
series of crossovers from under-damped to over-damped oscillations and to relaxation 
dynamics with increasing temperature are observed. 
This analysis is also relevant to understand the effect of ensemble of impurities
when a wide distribution of parameters has to be taken into account as in Ref.~\cite{shnirman05}.
In fact at any fixed temperature each impurity may display a specific dynamical behavior. 
As a consequence, various sets of impurities may effect the qubit dynamics in 
qualitatively different ways.
\begin{figure}[t!]
\centering
\includegraphics[width=0.35\textwidth]{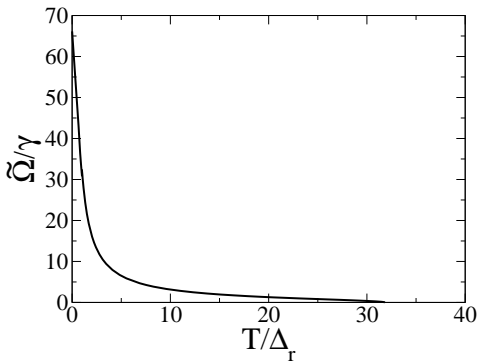}
\caption{Quality factor of the damped oscillations $Q(T)= \tilde \Omega(T)/\gamma(T)$
from Eqs.(\ref{eq:omega1}), (\ref{eq:gamma1}) and (\ref{eq:omega2}), (\ref{eq:gamma2}).
Parameters are fixed as in Fig.\ref{fig:omega}.
}
\label{fig:quality}
\end{figure}

Our goal is to establish to which extent the qubit may probe the variety of regimes of its
environment dynamics.
The important scale allowing the identification of Gaussian/non-Gaussian dynamical regimes
is extracted from the {\em equilibrium} auto-correlation function of the observable $\tau_z$ 
which directly couples to the qubit. We remark that the thermal initial state of the
Spin-Boson system, which is implied by the {\em equilibrium} correlation function, 
may originate peculiar time-dependencies.
Qualitatively different behaviors may in fact be displayed by the {\em non-equilibrium} correlation
function, which is evaluated for a factorized initial state of the Spin-Boson system~\cite{book,sassetti}.
Evaluation of {\em equilibrium} correlation functions for the Spin-Boson model is a non trivial
task. However it has been shown~\cite{book} that in the unbiased case $\varepsilon =0$, and
for small damping $K \ll 1$, the $\tau_z$ auto-correlation function does not depend on the initial correlated 
or factorized state.
Here we focus on this case and recall the characteristics of $S_\tau(\omega)$ more relevant for
our analysis, the interested reader may find details of the derivation in Refs.~\cite{book,sassetti}.
\begin{figure}[t!]
\centering
\includegraphics[width=0.30\textwidth]{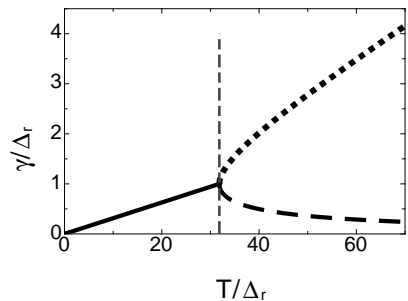}
\caption{
Dephasing rate $\gamma(T)$ from  Eqs.(\ref{eq:gamma1}) and (\ref{eq:gamma2}) (full line),
for temperatures $T < T^*(K)$ and relaxation rates $\gamma_{1/2}(T)$ from Eq.(\ref{eq:gamma12}) 
for $T \geq T^*(K)$. The rate $\gamma_1$ (dotted) increases with temperature, $\gamma_2$ (dashed)
shows the Kondo behavior. 
The value of $ T^*(K)/\Delta_r \approx 31.8$ is indicated by the dashed gray line.
Parameters are fixed as in Fig.\ref{fig:omega}
}
\label{fig:gamma}
\end{figure}

The crossover from under-damped to over-damped behavior with increasing temperature takes place  at 
\begin{equation}
T^*(K) \approx {\Delta_r \over \pi K k_B} \, ,
\label{eq:Tstar}
\end{equation}
where
$\Delta_r = \Delta (\Delta/\omega_c)^{K/(1-K)}$ is the 
renormalized tunneling amplitude in the Spin-Boson model.

The explicit form for $S_\tau(\omega)$ depends on temperature\cite{footnote1}

\noindent For $T < T^*(K)$ it reads
\begin{equation}
S_\tau (\omega)
={ \gamma + (\tilde\Omega - \omega) \tan\phi \over (\omega - \tilde\Omega)^2 + \gamma^2} 
+{ \gamma + (\tilde\Omega+ \omega) \tan\phi \over (\omega + \tilde\Omega)^2 + \gamma^2} \, ,
\label{eq:stau-low}
\end{equation}
where $\tan \phi = \gamma/\tilde\Omega$. 
The effective frequency $\tilde\Omega$ and the relaxation rate  are
\begin{eqnarray}
{\tilde \Omega(T) \over \Delta_r} &=&    
1 +  K \Big[ {\rm Re}{ \psi \Big(i {\Delta_r \over 2 \pi k_B T } \Big) }
	- \ln{\Big({\Delta_r \over 2 \pi k_B T }\Big)} - {\psi(1)}
      \Big] \nonumber \\
\label{eq:omega1}   \\
\gamma(T) 
&=& {S(\Delta_r) \over 4} \,,
\label{eq:gamma1}
\end{eqnarray}
for temperatures  $k_B T < \Delta_r$ and 
\begin{eqnarray}
{\tilde \Omega(T) \over \Delta_{T}} &=&  \sqrt{1 -  \Big({ T \over T^*} \Big)^{2-2K}}
\label{eq:omega2}
 \\ 
\gamma(T) &=& \pi K k_B T = {\Gamma \over 2} 
\label{eq:gamma2}
\end{eqnarray}
for larger temperatures $ \Delta_{r} \leq k_B T \leq k_B T^*(K)$.

Here $2 \Gamma$ is the white noise level $S(\omega) \approx 2 \Gamma$ 
for frequencies $\omega \ll 2 \pi / \beta $ in (\ref{eq:powerspec}).
Oscillators with frequencies  $2 \pi / \beta \leq \omega \ll \omega_c$
renormalize the tunneling amplitude to 
\begin{equation}
\Delta_T = \Delta_r (2 \pi k_B T/ \Delta_r)^K \, .
\end{equation}
In Fig.\ref{fig:omega} we show the overall behavior of  $\tilde \Omega(T)$. 
At very small temperatures $k_B T \ll \Delta_r $ (see inset in Fig.\ref{fig:omega}) 
$\tilde \Omega(T)$ increases $\propto T^2$. 
The two forms (\ref{eq:omega1}) and  (\ref{eq:omega2})  smoothly match on each other at $k_B T = \Delta_r$. 
A maximum is reached at $T= K^{1/2(1-K)}T^* > \Delta_r/k_B$, above this temperature $\tilde \Omega(T)$ 
decreases monotonously and approaches zero at $T^*$.  
Coherent oscillations are dephased on a scale $1/\gamma(T)$ which decreases monotonously
starting from $1/\gamma(0)= 2/(\pi K \Delta_r)$. The resulting quality factor of the
damped oscillations $Q(T) = \tilde\Omega(T)/ \gamma(T)$ is shown in Fig.\ref{fig:quality}.

For temperatures higher than $T^*$ the dynamics is incoherent and $S_\tau (\omega)$ has a different form
\begin{eqnarray}
S_\tau (\omega) &=&
2 \, {\gamma_2 \over \gamma_2  - \gamma_1 } { \gamma_1 \over \omega^2 + \gamma_1^2} 
\, + \,  2 \, {\gamma_1 \over \gamma_1  - \gamma_2 } { \gamma_2 \over \omega^2 + \gamma_2^2} 
\label{eq:stau-high}\\
\gamma_{1/2} &=& {\Gamma \over 2} \pm \sqrt{\Big({\Gamma \over 2}\Big)^2 - \Delta_T^2} \,.
\label{eq:gamma12}
\end{eqnarray}
Note that for $T \gg T^*(K)$ one of the two rates increases with temperature, $\gamma_1 \to \Gamma$, 
whereas the other shows the characteristic Kondo behavior, decreasing with temperature 
\begin{equation}
\gamma_2 \to {\Delta_T^2 \over \Gamma} ={\Delta_r \over K} ({\Delta_r \over 2 \pi k_B T})^{1-2K} \, .
\label{eq:kondo}
\end{equation}
These features are illustrated in Fig.\ref{fig:gamma}.

\begin{figure}[t!]
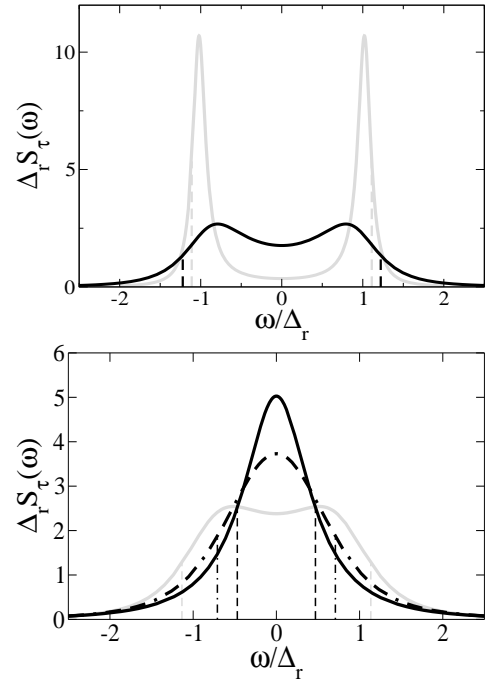

\centering
\includegraphics[width=0.35\textwidth,height=0.25\textwidth]{figure4top.eps}
\includegraphics[width=0.35\textwidth,height=0.25\textwidth]{figure4bot.eps}
\caption{Equilibrium correlation function $S_\tau(\omega)$ for increasing values of the temperature.  
Top:  from under-damped to over-damped regime and $T/\Delta_r= 3$ (light gray), 
$T/\Delta_r= 15$ (black). Dashed lines indicate the width $1/\tau_c$
at half height.
Bottom: from over-damped to  incoherent regime $T/\Delta_r=20 $ (light gray),
$T/\Delta_r=30 $ (dashed), $T/\Delta_r= 40$ (black).
Parameters are fixed as in Fig.\ref{fig:omega}.
}
\label{fig:corr-tauz}
\end{figure}

\begin{figure}[t!]
\centering
\includegraphics[width=0.35\textwidth]{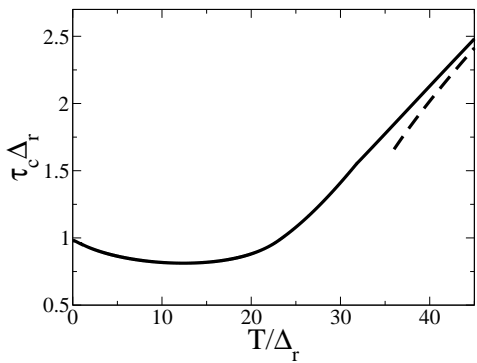}
\caption{Correlation time $\tau_c(T)$: for sufficiently small temperatures $\tau_c \approx 1/\Delta_r$.
For $T \gg T^*(K)$ the asymptotic behavior $ \approx 1/\gamma_2(T)$ is indicated (dashed).
The interpolating form for intermediate temperatures has been obtained from the condition 
$S_\tau(1/\tau_c) = S_\tau^{max}/2$.
Parameters are fixed as in Fig.\ref{fig:omega}.
}
\label{fig:tauc}
\end{figure}

The typical scale of the equilibrium fluctuactions of the Spin-Boson environment 
described by $S_\tau(\omega)$ defines the correlation time $\tau_c$.
Usually in the literature one is faced with  environment models characterized by dynamic 
fluctuations  which tend rapidly to zero with time. In these cases $\tau_c$ represents the 
order of magnitude of the width of the environment fluctuactions~\cite{kn:cohen}.
In the present case however looking at the Spin-Boson impurity as an environment characterized by $S_\tau(\omega)$
the identification of $\tau_c$ is not immediate due to the different forms of the equilibrium fluctuations
given by  Eqs.(\ref{eq:stau-low}) and (\ref{eq:stau-high}).
For high temperatures $S_\tau(\omega)$ is approximately a single Lorentzian centered at 
$\omega =0$ and width $\gamma_2$, leading to the identification $\tau_c \approx 1/\gamma_2(T)$, 
see Fig.\ref{fig:corr-tauz}.
In the opposite limit of very low temperatures $S_\tau(\omega)$ has a double peak structure
representing a bath responsible for oscillating fluctuations very weakly damped.
In this unusual environment regime, the typical scale of the impurity fluctuations is
represented by  $1/\tilde\Omega(T)$ which plays the role of $\tau_c$.
For intermediate temperatures in general two almost superimposed Lorentzians contribute to 
$S_\tau(\omega)$ and $\tau_c$ may be approximately identified from the condition 
$S_\tau(1/\tau_c) = S_\tau^{max}/2$. 
The resulting $\tau_c(T)$, illustrated in Fig.\ref{fig:tauc},
interpolates between the asymptotic behaviors at low and high temperatures. 
The slight reduction of $\tau_c(T)$ at intermediate temperatures is a consequence of the
crossover from under-damped to incoherent dynamics, as shown in Fig.\ref{fig:corr-tauz}.

Once the impurity correlation time is identified, the condition $v \tau_c(T)=1$ separates 
the weak-coupling regime of the qubit dynamics, from the strong coupling regime where non-Gaussian 
behavior shows up.
In the first case, when $ v \tau_c(T) \ll 1$, the standard master equation predicts exponential 
decay with the Golden Rule rate ${1 \over T_2}$ given in Eq.(\ref{dephasing}).
>From the above analysis  we expect the master equation
result to be valid in the following regimes:
For temperatures $T < T^*$ if $v/\tilde\Omega(T) \approx v/ \Delta_r \ll 1$, for
larger temperatures $T > T^*$ if $v / \gamma_2(T) \approx v \Gamma /\Delta_T^2 \ll 1$, this
condition can be cast in the following form 
\begin{equation}
{v \over \Delta_r} \ll \Big({T^* \over 2T}\Big)^{1-2K} \, .
\label{eq:crossoverT}
\end{equation}
Therefore, for small values of $v/\Delta_r$ a crossover from weak to strong coupling is
expected with increasing temperature. 
For the ensuing discussion here we report the expected value of the pure dephasing 
rate $1/T_2^*=  (E_C / E)^2 v^2 S_\tau(0)/ 2$ with Eqs.(\ref{eq:stau-low}) and (\ref{eq:stau-high}) 
\begin{eqnarray}
{1 \over T_2^*} &=&   \Big({E_C \over E}\Big)^2  {4 v^2 \gamma(T) \over \gamma(T)^2 + \tilde\Omega(T)^2} 
 \qquad  \qquad T<T^*
\label{eq:GR1}
\\
{1 \over T_2^*} &= & \Big({E_C \over E}\Big)^2 \, \Big({ v \over \Delta_T}\Big)^2 \Gamma 
 \qquad  \qquad  \qquad T>T^*
\label{eq:GR2}
\end{eqnarray}
The two forms match on each other at $T^*$, and it is easy to show that Eq.(\ref{eq:GR2})
approximates $1/T_2^*$ also for $T \ll T^*$.

\section{Qubit dynamics: path-integral approach}
\label{sec:qubit}

\noindent 
The discussion of the previous Section has evidenced the existence of a large
parameter regime where the Gaussian approximation of the Ohmic Spin-Boson model
does not apply. 
In this Section we study the qubit dynamics via a path-integral approach 
which includes as a special case the regime of Gaussian behavior of
the impurity dynamics. We find exact expressions which we discuss for
finite temperatures, specifically for $k_B T \geq \sqrt{\Delta_r^2 + v^2}$.

We focus on the so called pure-dephasing regime, $E_J=0$, which represents the point
of maximum noise sensitivity of the qubit. Thus the more interesting in the perspective 
of using the qubit as a ``noise'' analyzer. 
In the pure dephasing regime the charge on the qubit island is a constant
of motion since $[\mathcal{H}, \sigma_z]=0$, dephasing
being described by the decay of $\langle \sigma_{x/y} \rangle$ or equivalently of
the coherences $\langle \sigma_{\pm} \rangle$.
A simple analysis shows that the coherences are related to correlation functions involving the Spin-Boson 
variables, specifically we found \cite{PRL02,chemphys04} 
\begin{figure}[t!]
\centering
\includegraphics[width=0.35\textwidth,height=0.25\textwidth]{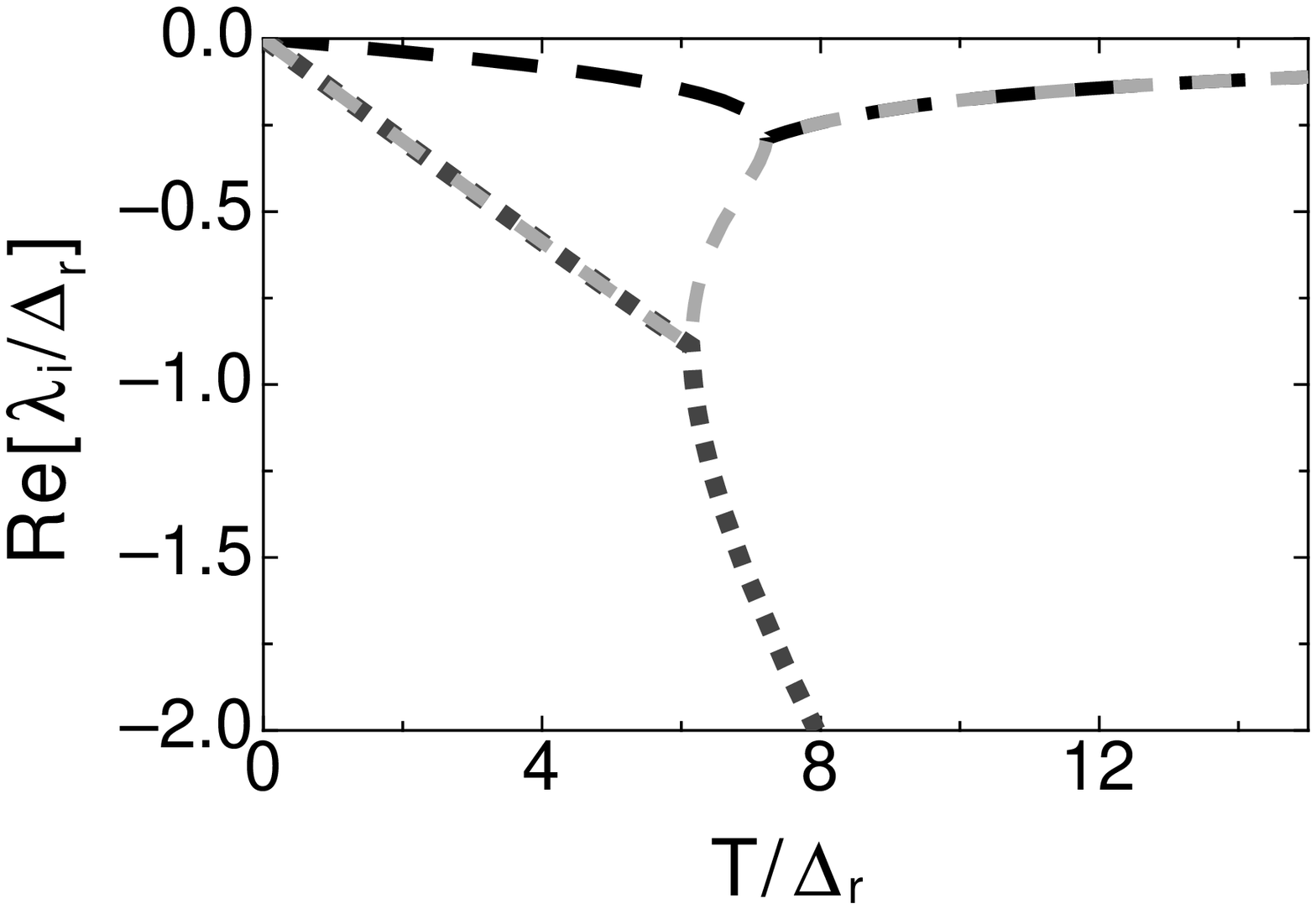}
\includegraphics[width=0.35\textwidth,height=0.25\textwidth]{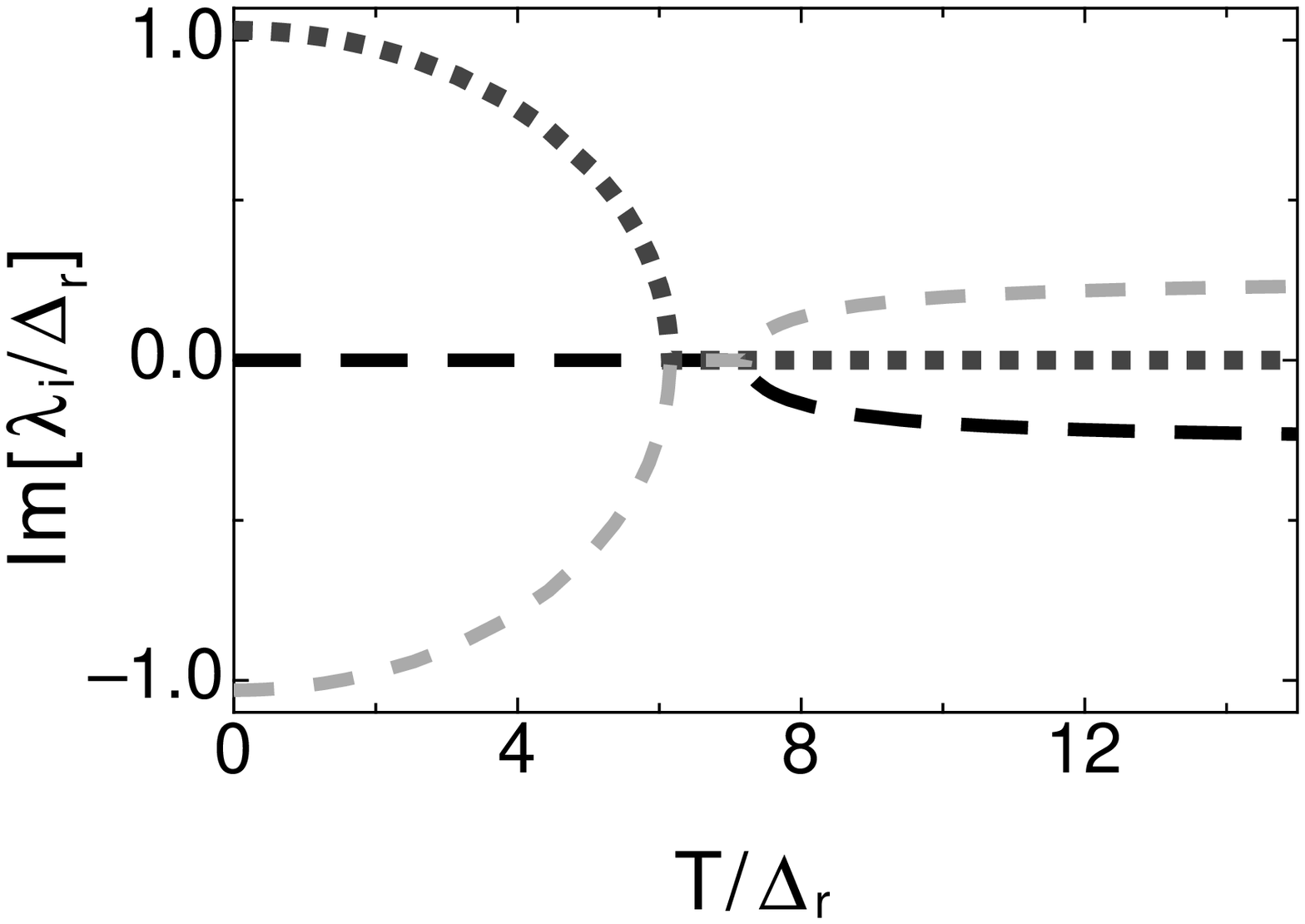}
\caption{Top: real parts of the exact solutions $\lambda_i$ of the pole equation $D(\lambda)=0$ 
as a function of temperature for $k_B T \geq \sqrt{v^2+\Delta_r^2} \approx 0.87 \Delta_r$. 
Bottom: corresponding imaginary parts.
Two $\lambda_i$ are complex conjugate below $T_- \approx 6.1 \, \Delta_r$ and above $T_+ \approx 7.2 \, \Delta_r$.
All $\lambda_i$ are real at intermediate temperatures $T_- < T < T_+$. 
The dominant pole is  real (black dashed) until $T_+$ where the
character of the dominant solution changes. For $T > T_+$  the dominant poles are
complex conjugate. Parameters are  $v/\Delta_r =0.25$, $\omega_c/\Delta_r=30$ and $K=0.05$.}
\label{fig:smallv}
\end{figure}
\begin{eqnarray}
{\langle \sigma_-(t) \rangle \over \langle \sigma_-(0) \rangle } &= &
e^{i E_C t} \, 
{\rm Tr}_{SB} \big\{ e^{-i \mathcal H_{SB-} t} \, 
\rho_{\tau}(0) \otimes w_{\beta} \, 
e^{i \mathcal H_{SB+}t} \big\}\, \nonumber \\
& \equiv&  
\, e^{i E_C t} \, C_{-+}(t)
\label{element}
\end{eqnarray}
where we have chosen a  factorized initial density matrix for the qubit-impurity, 
$\rho(0) =  \rho_{\sigma}(0) \otimes \rho_{\tau}(0)$, with the
impurity $\tau$  initialized in the mixed state 
$\rho_\tau(0) = \frac{1}{2} \, \hat I \, + 
\,\frac{1}{2} \delta p(0) \,\tau_z$ and the  bosonic bath in its thermal equilibrium state 
 $w_{\beta}$.
The two conditional impurity Hamiltonians $\mathcal H_{SB\pm}$ depend on the qubit state and read
$\mathcal H_{SB\pm}= \mathcal H_{\rm SB} \pm \frac{v}{2} \tau_z$.

In Ref.~\cite{chemphys04} it has been shown that the Laplace transform of 
the correlator $C_{-+}(t)$ reads
\begin{eqnarray}
{\widehat C_{-+}(\lambda)}& = & \frac{1}{D(\lambda)} \, \left [ \, 
\lambda + K_1(\lambda) 
- i v \delta p(0) \, \right ] 
\label{eq:C-exact}
\\
D(\lambda) &=& \lambda^2 +  v^2 
+ \lambda K_1(\lambda) +
i v K_2(\lambda) \, .
\label{eq:poles-exact}
\end{eqnarray}  
\begin{figure}[t!]
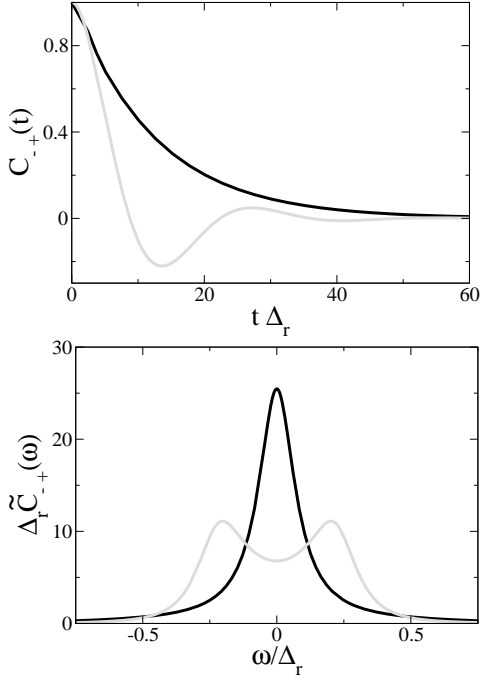

\centering
\includegraphics[width=0.35\textwidth,height=0.25\textwidth]{figure7top.eps}
\includegraphics[width=0.35\textwidth,height=0.25\textwidth]{figure7bot.eps}
\caption{Top: $C_{-+}(t)$ for $v/\Delta_r=0.25$ and $k_B T/\Delta_r=4$ (black), $k_B T/\Delta_r=15 $ 
(gray). Bottom: corresponding Fourier transform.
Here $K=0.05$, $\omega_c/\Delta_r=30$, $\delta p(0)=0$.
}
\label{fig:csmallv}
\end{figure}

An exact formal series expression in $\Delta$ for the kernels $K_1(\lambda)$, $K_2(\lambda)$ 
has been derived in Ref.\cite{chemphys04}.  
In the Markovian regime for the harmonic bath, i.e. for $K \ll 1$ and temperatures 
$\sqrt{\Delta_r^2 + v^2} \leq  k_B T \ll \omega_c$, 
all contributions to $K_1(\lambda)$ and $K_2(\lambda)$ 
of order higher  than $\Delta^2$ cancel out exactly.
The lowest order contributions \cite{footnote2}
do not depend on the coupling $v$ and coincide with the kernels entering the
dynamics of $\vec{\tau}$ in the uncoupled case ($v=0$)~\cite{book} 
which read
\begin{eqnarray}
\label{NIBA1L}
{\mathcal K}_1(\lambda) &=& 
 \Delta_T^2 \,
{\lambda + \Gamma  \over \epsilon^2 + (\lambda+ \Gamma )^2} \\
\label{NIBA2L}
{\mathcal K}_2(\lambda) &=& -
 \pi K \Delta_T^2 \, {\epsilon \over \epsilon^2 + (\lambda+ \Gamma )^2
} \, .
\end{eqnarray}
Inserting Eqs.(\ref{NIBA1L})~-~(\ref{NIBA2L}) in (\ref{eq:C-exact}) and (\ref{eq:poles-exact}) 
${\widehat C_{-+}(\lambda)}$ is readily found as 
\begin{eqnarray}
\begin{array}{ll}
&\!\!\!\!\!\!{\widehat C_{-+}(\lambda)} =  {[(\lambda+ \Gamma )^2 +  \epsilon^2] \,
[\lambda - i v \delta p(0)] + \Delta_T^2 (\lambda+ \Gamma )\over D(\lambda)} 
\label{eq:C-laplace}
\\
&\!\!\!\!\!\!D(\lambda) = 
(\lambda^2 +  v^2) \, [(\lambda+ \Gamma )^2 +  \epsilon^2] 
+ \Delta_T^2 \lambda  (\lambda+ \Gamma ) \\
&\qquad- i \pi K v \epsilon \Delta_T^2  \, .
\end{array}
\label{denominator}
\end{eqnarray}

The scales entering the time evolution of $C_{-+}(t)$ are found from the solution
of the pole equation $D(\lambda)=0$, which have been reported in Ref.\cite{chemphys06}.
Here we specify to the unbiased case $\epsilon=0$, the goal being to elucidate the correspondence
with the expected Gaussian/non-Gaussian dynamical regimes as deduced from the equilibrium
correlation function $S_\tau(\omega)$ discussed in Section \ref{sec:impurity}. 
\begin{figure}[t!]
\centering
\includegraphics[width=0.35\textwidth,height=0.25\textwidth]{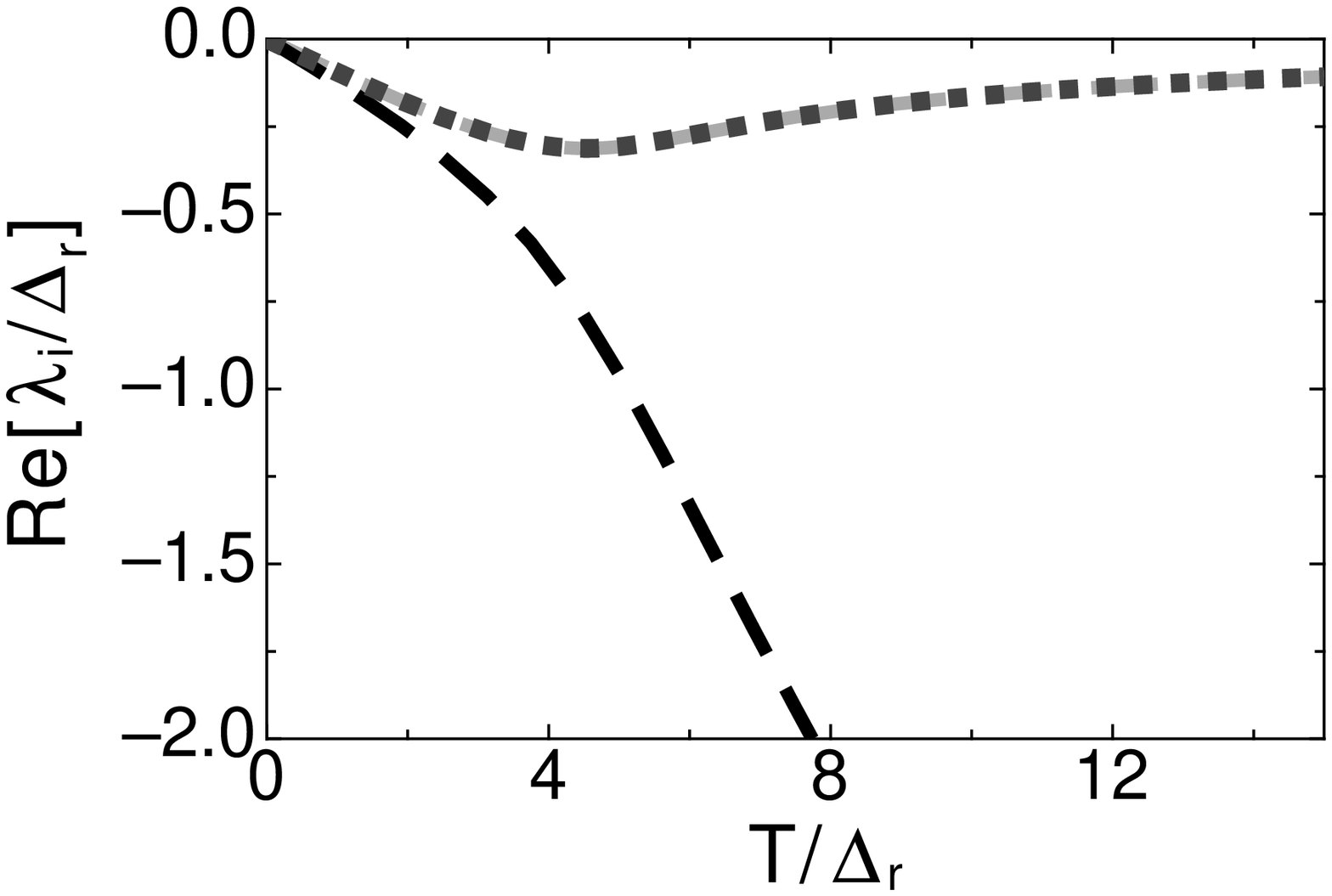}
\includegraphics[width=0.35\textwidth,height=0.25\textwidth]{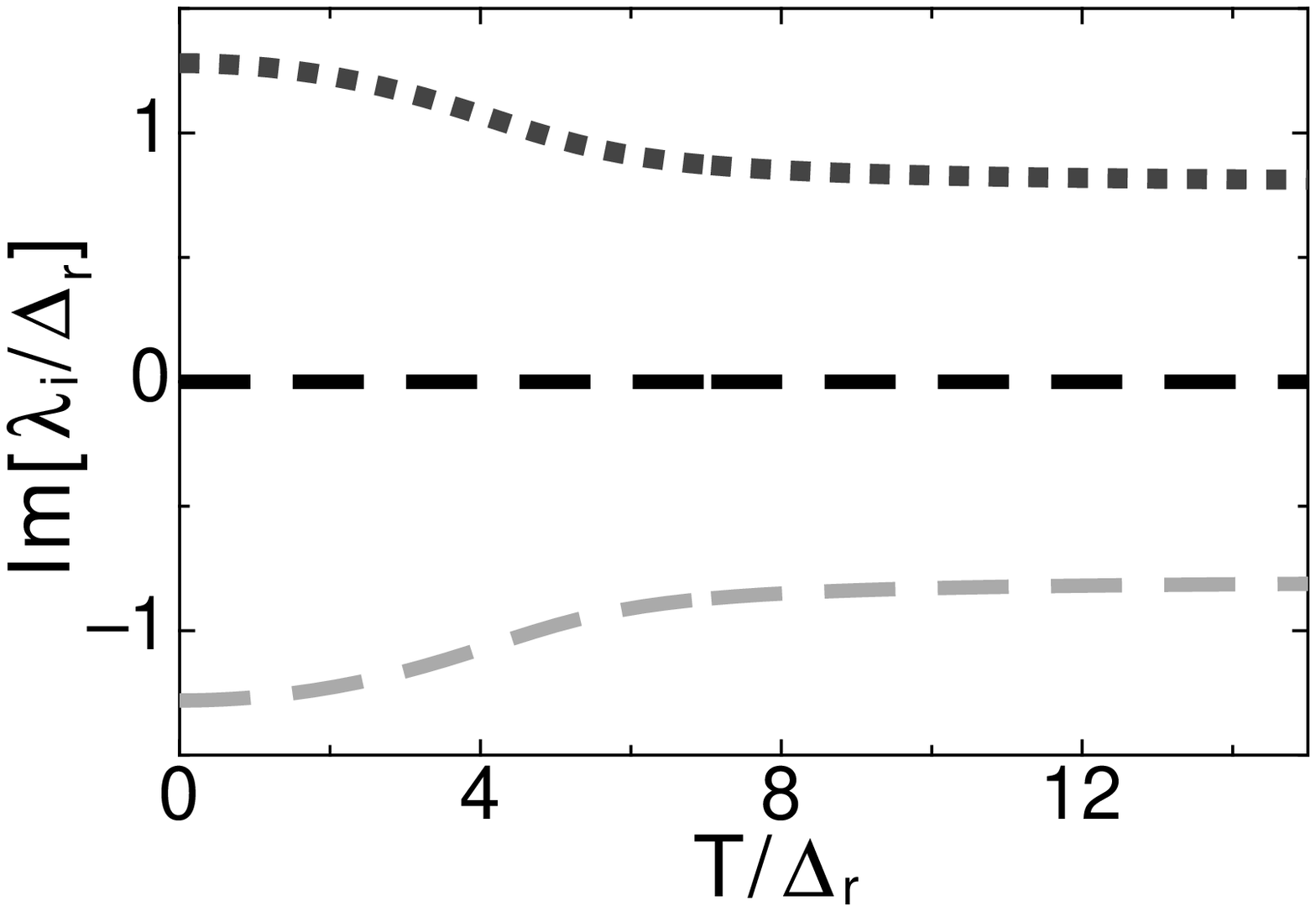}
\caption{Top: real parts of the exact solutions $\lambda_i$ of the pole equation $D(\lambda)=0$ 
as a function of temperature for  $k_B T \geq \sqrt{v^2+\Delta_r^2} \approx 1.2 \Delta_r$. 
Bottom: corresponding imaginary parts.
Two $\lambda_i$ are complex conjugate and one is real for any temperature.
The dominant poles are always complex conjugate. 
In this regime no crossing takes place among the ${\mathcal R}e[\lambda_i]$ and
the dominant root is non-monotonic with $T$.
Parameters are  $v/\Delta_r =0.8$,  $K=0.05$, $\omega_c/\Delta_r=30$.
}
\label{fig:intv}
\end{figure}

\subsection{Pure dephasing due to a unbiased impurity}

When $\epsilon=0$ the pole condition  $D(\lambda)=0$ with  Eq.(\ref{denominator})  
reduces to a cubic equation
which has either one real and two complex conjugate solutions, or three real solutions.
We denote the three roots as $\lambda_0 = - \Lambda_0 \in {\mathcal R}e$ and
$\lambda_{1/2} = - \Lambda \pm i \delta E$, where $\delta E$ is either real or purely imaginary.
The expression of $C_{-+}(t)$ in terms of the $\lambda_i$ is obtained by
inverting the Laplace transform (\ref{eq:C-laplace}) and reads

\begin{eqnarray}
&& \begin{array}{ll}
C_{-+}(t) & = A \, e^{- \Lambda_0 t} \,+\, (1-A)\, \cos{(\delta E t)} \, e^{- \Lambda t}+ \\
	  & +  B \, \sin{(\delta E t)} \, e^{- \Lambda t}
\end{array} \\
&& 
A \,=\, {-2 \Lambda_0 \Lambda + \Delta_T^2 - i 2 \delta p(0) \, v \Lambda   
			\over (\Lambda - \Lambda_0)^2 + \delta E^2} \\
&& 
B \,=\,{ \big[  \Lambda (1-A) \,+ \, \Lambda_0 A \,- i  \delta p(0) \, v \big] \over \delta E}  \, .
\label{eq:ct}
\end{eqnarray}
It is possible to show that the character of the roots depends on $v/\Delta_T$ and on
the temperature.
In particular for  $v < \Delta_T/ 2 \sqrt{2}$ we can identify two temperatures
\begin{eqnarray}
k_B \, T_\pm &\approx& {1 \over \pi K} 
\Big[- \Big({v \over 2}\Big)^2 + {5 \Delta^2 \over 8} + {\Delta^4 \over 32 v^2} \nonumber \\
&\pm &{\sqrt{\Delta^8 -24 v^2 \Delta^6 +192 v^4 \Delta^4 - 512 v^6 \Delta^2} \over 32 v^2}
\Big]^{1/2}
\end{eqnarray}
such that for $T < T_-$ and $T > T_+$ one solution  is real and two are complex conjugate,
whereas for intermediate temperatures the three solutions are real.
For $v > \Delta_T/ 2 \sqrt{2}$ there is always one real and two complex conjugate solutions.
Analytic expressions for the roots are quite cumbersome, approximate forms
have been reported in~\cite{chemphys04}. 
Here we discuss the physically relevant regimes where crossover are expected  in the behavior of 
$C_{-+}(t)$.
To this end we focus on the dominant $\lambda_i$, i.e. the smallest in absolute value.
The analysis is conveniently performed distinguishing regimes where $v/\Delta_T <1$ or
$v/\Delta_T >1$

\begin{figure}[t!]
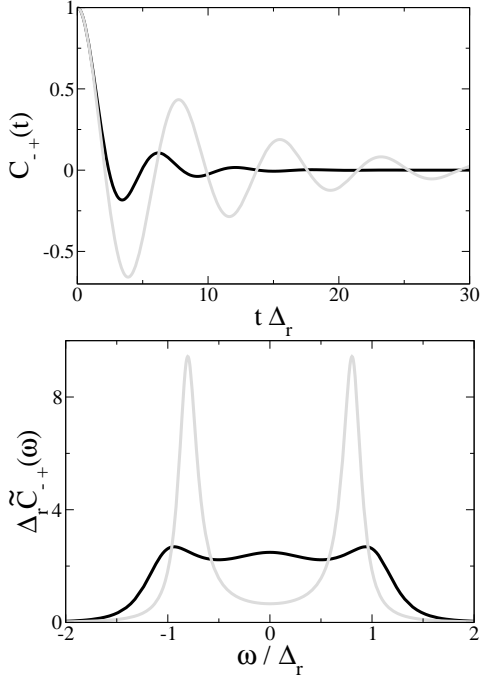

\centering
\includegraphics[width=0.35\textwidth,,height=0.25\textwidth]{figure9top.eps}
\includegraphics[width=0.35\textwidth,,height=0.25\textwidth]{figure9bot.eps}
\caption{Top: $C_{-+}(t)$ for $v/\Delta_r=0.8$ and $k_B T/\Delta_r=4 $ (black),
$k_B T/\Delta_r=15 $ (gray). Bottom: corresponding Fourier transform.
Here $K=0.05$, $\omega_c/\Delta_r=30$, $\delta p(0)=0$.
}
\label{fig:cintv}
\end{figure}

{\em Case  $v/\Delta_T \ll 1$}
In this regime the characteristics of the dominant pole change qualitatively with increasing
temperatures (with lower bound $k_B T > \sqrt{v^2 + \Delta_r^2}$).
For temperatures low enough to fulfill the condition
\begin{equation}
{k_B T \over \Delta_T} \ll{1 \over 4 \pi K v} \approx {k_B \, T_+ \over \Delta_r}
\label{eq:condition}
\end{equation}
the dominant scale is real and reads $\Lambda_0 \approx  ({v \over \Delta_T} )^2 \Gamma$.
In (\ref{eq:ct}) $A \approx 1 -(\Gamma /2 \Delta_T)^2$ and $B \approx - i \delta p(0) v / \Delta_T$ 
therefore
\begin{equation}
C_{+-}(t) \approx \exp{[- \Lambda_0 t]} \, .
\label{eq:gauss}
\end{equation} 
With increasing temperature, above $T_+$ the dominant scales are complex conjugate,
$\lambda_{1/2} = - \Lambda \pm i \delta E \approx - \Delta_T^2 / 2 \Gamma \pm i v$. Figure \ref{fig:smallv} illustrates
the crossover between the two regimes for $v/\Delta_r=0.25$.
Since $A \ll v /\Gamma$ and $B \approx -i \delta p(0)$, we get
\begin{equation}
C_{+-}(t) \approx  \cos{(v t)} \, e^{- \Lambda t} \, -\, i \delta p(0) \sin{(v t)} \, e^{- \Lambda t} \,.
\label{ct-nongauss}
\end{equation}
As expected, for small enough temperature the Gaussian approximation for the structured bath applies
and a single scale dominates the qubit dynamics.
It is easily seen that  the condition (\ref{eq:condition})
corresponds to (\ref{eq:crossoverT}) which was derived from the weak coupling
criterion $v \tau_c \ll 1$. Moreover $\Lambda_0 = 1/T_2^*= v^2 S_\tau(0)/2$ as given in
Eq.(\ref{eq:GR2}). 
For higher temperature non Gaussian effects show up. The qualitative change is characterized by 
damped oscillations at frequency $\approx v$ in $C_{+-}(t)$  as shown in Fig.\ref{fig:csmallv}, and
beatings at frequencies $E_C \pm v$ in the coherences Eq.(\ref{element}).   
Note that the coupling strength $v$ only enters the induced frequency shift, whereas the decay 
rate shows the Kondo behavior $\Lambda \propto T^{2K -1}$, cfr Eq.(\ref{eq:kondo}). \\

{\em Case  $v/\Delta_T \gg 1$}
For larger values of $v/\Delta_T$ the system stays in the regime where the dominant scales are complex
conjugate and show a nontrivial temperature dependence. 
At temperatures $T \geq \sqrt{v^2 +\Delta_r^2}$, the two poles read 
$\lambda_{1/2} \approx - (\Delta_T/v)^2 \Gamma/2 \pm i v$,
with increasing $T$ instead $\lambda_{1/2} \approx - \Delta_T^2 /( 2 \Gamma) \pm iv$.
The qubit dynamics follows Eq.(\ref{ct-nongauss}).
The dominant rate $\Lambda$ is non-monotonous first increasing with temperature and than decreases
$\propto T^{2K-1}$.
For large enough temperatures the decay of the coherences does not depend on $v$, as expected.

This qualitative behavior is already present for intermediate values of $v/\Delta$,
as shown in Fig.\ref{fig:intv}. The poles $\lambda_i$ never cross, therefore there is
no change in the character of the qubit dynamics. In the specific case considered 
at small/intermediate temperatures the  real parts of the three  poles are of the same order.
This is reflected in the  Fourier transform of  $C_{+-}(t)$ where three Lorentzians 
can be identified,
one centered at $\omega=0$, the others at $\approx \pm v$, see Fig.\ref{fig:cintv}.

\section{Discussion}
\label{sec:discussion}

The presented analysis has shown that a damped impurity may behave as an
effective short-time correlated or as a non-Gaussian environment depending
both on its coupling with the qubit and on temperature. The relevant scale 
separating the two regimes is given by $\tau_c$. 
In  Section \ref{sec:impurity} the correlation time of the 
unbiased Spin-Boson model has been found for small damping $K\ll1$
at any temperature.
We have shown that the qubit may act as a detector  of
non-Gaussian dynamical behavior, the most evident effect being the
occurrence of beatings which are expected with increasing 
temperature. We remark that the results we have illustrated have been
derived within the NIBA which limits temperatures to values larger than $\sqrt{v^2+\Delta_r^2}$. 
A interesting issue is to analyse the small temperature regime
where we expect that crossover effects may take place also for large values
of $v/\Delta_T$.
A detailed analysis of the low temperature regime 
 can be performed within the systematic weak damping approximation~\cite{book} and will be 
reported elsewhere \cite{short2007}.


\begin{thebibliography}{10}
\bibitem{footnote1} The above analytic forms have 
been derived in  Ref.~\cite{book} within the so called
Non-Interacting-Blip-Approximation (NIBA) which is exact in the considered regime 
($K\ll 1$ and $\epsilon=0$).
\bibitem{footnote2} In the Spin-Boson literature the approximation of the kernels to order
$\Delta^2$ is referred to as Noninteracting-Blip-Approximation (NIBA). 
It's validity regimes include the considered Markovian case 
$K \ll 1$ and $k_B T \geq \sqrt{v^2 + \Delta_r^2}$~\cite{book}.
\end{thebibliography}
\end{document}